\documentclass[aps,prl,reprint,twocolumn]{revtex4}

\usepackage{newtxtext}
\usepackage{newtxmath}
\usepackage[T1]{fontenc}

\usepackage{xcolor}
\usepackage{epsfig}
\usepackage{graphicx}
\usepackage{amsmath}
\usepackage{natbib}
\usepackage{braket}

\begin{document}
\title{Catability as a metric for evaluating superposed coherent states}
    \author{\v{S}imon Br\"{a}uer, Jan Provazn\'{i}k, Vojt\v{e}ch Kala, Petr Marek}
\date{\today}
\affiliation{Optics Department, Faculty of Science, Palack\'{y} University, 17. Listopadu 12,
 77146 Olomouc, Czech Republic}

\begin{abstract}
Superposed coherent states are central to quantum technologies, yet their reliable identification remains a challenge, especially in noisy or resource-constrained settings. We introduce a novel, directly measurable criterion for detecting cat-like features in quantum states. The criterion is based on the concept of nonlinear squeezing and bypasses the need for full state tomography of the tested states. The numerical results confirm its robustness under loss and its potential for experimental implementation. The method naturally generalizes to more exotic superpositions, including multi-headed cat states.
\end{abstract}

\maketitle

Superposed coherent states, also known as coherent cat states, are inspired by the famous Schr\"{o}dinger thought experiment~\cite{schrodinger1935}. Apart from being particularly useful for discussing the foundations of quantum mechanics, they play a crucial role in practical applications; coherent cat states can be used to encode information in quantum computation~\cite{Jeong2002Mar, ralph2003,lund2008,Lee2013Feb, Omkar2020Aug, Omkar2021Mar, hastrup2022,lee2024,su2022,leghtas2013}, as highly sensitive probes in quantum metrology~\cite{duivenvoorden2017,pan2025}, and as fundamental resources in breeding protocols for the experimental preparation of more complex quantum states~\cite{lund2004,sychev2017,weigand2018,hastrup2022}. Due to their versatility, cat states are the sought-after target of many experiments; to this day, cat states of various quality have been prepared in traveling light~\cite{jezek2012,yukawa2013,eaton2022,endo2025}, microwave resonators \cite{vlastakis2013,ofek2016,he2023}, trapped ions~\cite{monroe1996,kienzler2016}, atoms~\cite{omran2019},  or hybrid systems~\cite{hacker2019}, and there are several proposals for their preparation in optomechanical systems~\cite{shomroni2020,hauer2023}.

One of the central challenges in all related experiments is the characterization of the prepared quantum states. A commonly considered indicator is non-Gaussianity of the state~\cite{filip2011,walschaers2021,lachman2022}, commonly defined as deviation from a Gaussian shape of its Wigner function. Although there are several methods for its detection~\cite{Lee2011May,lachman2022,fiurasek2013,park2015, Kwon2017Apr}, they offer only a limited view of the nature of the state. The conventional approach used in practice utilizes fidelity~\cite{sychev2017,bild2023,he2023}, defined as the geometric overlap between two quantum states, to provide a more complete, global comparison of the experimentally produced state and its theoretical counterpart. Fidelity is appealing due to its intuitive interpretation, but it lacks nuance when applied to states occupying infinite-dimensional Hilbert spaces. Its value can be affected in fundamentally different ways that are difficult to distinguish from fidelity alone. For example, low fidelity might indicate that the tested state is qualitatively different; equally it might also be a mixture of qualitatively similar states. Furthermore, fidelity typically requires full knowledge of the quantum state, which must be obtained through quantum state reconstruction~\cite{yukawa2013,eaton2022,endo2025,monroe1996,kienzler2016,vlastakis2013,ofek2016,he2023,genoni2013}.

A directly measurable alternative to fidelity is available in some quantum states that manifest reduced fluctuations of some observable operator. The original squeezed states~\cite{stoler1970,lu1971,yuen1976} displayed reduced fluctuations in the canonical quadrature operators of harmonic oscillators. These states became indispensable for the preparation and manipulation of quantum states~\cite{yukawa2013,eaton2022,endo2025,braunstein2005,yadin2018,hamilton2017}, with applications in quantum metrology~\cite{jia2024} and quantum communication protocols~\cite{gottesman2001,oruganti2025}. When it comes to state characterization, the concept of squeezing is beneficial because it requires only a single measurement basis. While this omits some information about the state, it is often sufficient as the performance of many applications depends only on the squeezed variance of the quadrature operators~\cite{jia2024,park2022,braunstein1998,furusawa1998}.

The concept of squeezing can be broadened to nonlinear squeezing, where an arbitrary observable operator exhibits reduced fluctuations~\cite{brauer2025,kala2025}. Whereas traditional squeezing principally represents a similarity to some eigenstate of a canonical quadrature operator, nonlinear squeezing characterizes correspondence to some eigenstate of the general observable. In principle, nonlinear squeezing can quantify closeness to any quantum state, which can be identified as an eigenstate of some observable. This extension can be advantageously used to evaluate the quality of complex non-Gaussian quantum states used as resources in quantum technologies, for example, those displaying cubic or quartic nonlinearity~\cite{miyata2016,konno2021,brauer2021,kala2022}, or the Gottesman-Kitaev-Preskill states~\cite{marek2024}. 

In this letter, we identify such an operator for superposed coherent states and use it to define catability: a directly measurable quantity that can be used to identify and evaluate cat states, clearly showing their non-Gaussian nature. We compare its sensitivity to non-Gaussian features for decohered cat states with fidelity. We then show that its evaluation requires only three sets of number operator measurements instead of full quantum tomography. We further demonstrate that it can be generalized to a broader class of cat states.

The balanced coherent cat states are defined as
    \begin{equation}\label{eq:cat}
        \ket{\alpha,\pm} = \frac{1}{\sqrt{2}} \frac{\left( \ket{\alpha} \pm\ket{-\alpha} \right)}{\sqrt{1\pm \exp (-2|\alpha|^2)}}
    \end{equation}
and represent the ground eigenstates of a class of directly observable positive semi-definite operators
\begin{equation}\label{eq:O_for_cat}
        \hat{\mathcal{O}}_{(\pm)}(\alpha,\gamma) = \left( \hat{a}^{\dagger2} - \alpha^{*2} \right)\left( \hat{a}^{2} - \alpha^{2} \right) + \gamma\left( 1 \mp \hat{\Pi} \right) \;.
\end{equation} 
The symbols $\hat{a}$ and $\hat{a}^{\dag}$ are the annihilation and creation operators of the harmonic oscillator, and $\hat{\Pi}$ embodies the parity operator. The first term tests the separation of the coherent peaks, while the second term uses the parity operator to verify the presence of quantum coherence. The real parameter ${\gamma > 0}$ controls the importance of achieving either of these features. Regardless of its value, the ideal cat states~\eqref{eq:cat} are the eigenstates of the positive semi-definite operators~\eqref{eq:O_for_cat} corresponding to the lowest eigenvalue ${\lambda_{0} = 0}$.
The parameter ${\gamma>0}$ in \eqref{eq:O_for_cat} represents a tunable trade-off between the two ingredients of the witness: the peak separation and the parity terms. In practice, $\gamma$ balances sensitivity to branch separation versus quantum coherence. Ideal cat states remain ground states of ${\hat{\mathcal O}_{(\pm)}(\alpha,\gamma)}$ for any $\gamma>0$, and the definition of catability \eqref{eq:catability} minimizes over $\gamma$ so that the optimal weight is chosen to suit the data. For very large $\gamma$, the criterion becomes primarily sensitive to parity, while for small $\gamma$ it favors peak separation. In experiments, moderate $\gamma$ values naturally balance both features. In our subsequent numerical analysis, we restrict ${\gamma \in [0,5]}$ for convenience. In principle, the optimization is not limited to this range.

These operators can be used to define the catability of an arbitrary quantum state  $\hat{\rho}$ as
 \begin{equation}\label{eq:catability}
        \xi_{(\pm)}(\alpha) = \min_{\gamma} \frac{ 
            \text{Tr}\left(  \hat{\mathcal{O}}_{(\pm)}(\alpha,\gamma) \hat{\rho} \right) 
        }{
            \min_{\hat{\rho}_{G}} \text{Tr}\left( \hat{\mathcal{O}}_{(\pm)}(\alpha,\gamma) \hat{\rho}_{G}  \right)
        },
    \end{equation}
where the minimization in the denominator is performed over the set of all Gaussian states \cite{SM}. Note that the optimization over Gaussian states needs to be performed only once. The normalization can be quickly determined using a precomputed lookup table~\cite{source}. The whole expression is minimized over all the possible ${\gamma > 0}$ parameters. The numerator can be interpreted as nonlinear squeezing on its own; it is the second moment of the operator 
\begin{equation}
    \sqrt{\hat{\mathcal{O}}_{(\pm)(\alpha,\gamma)}}
\end{equation}
and even becomes equal to its variance when the first moment is set to zero, similarly to the approach presented in~\cite{marek2024}. The nonlinear squeezing can be normalized with the best value attainable by Gaussian states, making the following practical interpretation possible.
    \begin{enumerate}
        \item If $\xi_{(\pm)}(\alpha) = 0$, then $\hat{\rho}$ is an eigenstate of the operator $\hat{\mathcal{O}}_{(\pm)}(\alpha,\gamma)$ and represents a perfect cat state $\ket{\alpha,\pm} $, which is necessarily non-Gaussian.
        \item If $0 < \xi_{(\pm)}(\alpha) < 1$, then $\hat{\rho}$ is an imperfect, approximate cat state with amplitude $\alpha$, but we can still confirm its non-Gaussian nature.
        \item If $\xi_{(\pm)}(\alpha) \geq 1$, we cannot say. It might be the correct cat state with too much decoherence, a different cat state, or a different state altogether. 
    \end{enumerate}
Imperfect or approximate cat states are understood here as realistic quantum states that may be expressed as superpositions or mixtures of ideal cat states and other components, arising, for example, from loss, decoherence, and other experimental imperfections~\cite{SM}.

Catability allows us to determine the quality of an experimental cat state with specific parity and amplitude~$\alpha$. It can also serve as a straightforward witness indicating whether an unknown state approaches the desired cat state. We can also define a global catability 
\begin{equation}
    \xi = \min_{\alpha,\pm} \xi_{(\pm)}(\alpha)
\end{equation}
to check whether a given unknown test state is a non-Gaussian approximation of some undetermined cat state. 

Let us now demonstrate the application of catability in some realistic scenarios. We consider ideal cat states~\eqref{eq:cat} decohered by pure loss; this is a common occurrence in optical experiments and quickly reduces the quantum non-Gaussian properties of the affected states~\cite{lejeannic2018,provaznik2024}. Pure loss can be modeled with an unbalanced beam splitter where the transmitted signal interacts with the environment in a vacuum state. Its amplitude transmissivity ${0 \leq \eta \leq 1}$ determines the amount of incurred loss. The output state can be computed with Kraus operators~\cite{ivan2011}
\begin{equation}
    \begin{gathered}
        \hat{\rho}_{\text{out}} = \sum_{n=0}^{\infty} \hat{M}_{k} \hat{\rho} \hat{M}_{k}^\dagger \\
        \;\;\text{where}\;\;
        \hat{M}_{k} = \sqrt{ \frac{(1 - \eta^2)^k}{k!} } \, \eta^{\hat{a}^\dagger \hat{a}} \, \hat{a}^{k}
        \;\text{.}
    \end{gathered}
\end{equation}
Losing half of the signal is sufficient to remove all negativity from the Wigner functions of any non-Gaussian state~\cite{mari2012}.

We also consider approximate cat states that can be prepared from finite superpositions of photons using Gaussian squeezing ${\hat{S}(r) = \exp[\frac{r}{2} (\hat{a}^{2} - \hat{a}^{\dagger2})]}$. The states
\begin{gather}
    \label{eq:singlephoton-approx}
    \hat{S}(r)\ket{1} \quad\text{and} \\
    \label{eq:even-approx}
    \hat{S}(r)\left(\sqrt{\omega}\ket{0} + \sqrt{1-\omega}\ket{2}\right)
\end{gather}
approximate small cats with even and odd parities, respectively. These approximations are also highly susceptible to loss and appear in experiments with traveling light, where the cat states are usually generated using generalizations of photon subtraction and photon number counting techniques~\cite{takahashi2008,lejeannic2018,takase2021,yukawa2013,eaton2022,endo2025}.

We evaluate the catability for these states and compare it with the well-established fidelity. To facilitate direct comparison, we introduce the normalized infidelity of a quantum state $\hat{\rho}$ with the target cat state \eqref{eq:cat} as
\begin{equation}\label{eq:normed_infi}
    \zeta_{(\pm)}(\alpha) = \frac{
        1 - \langle \alpha,\pm|\hat{\rho}|\alpha,\pm\rangle
    }{
        \min_{\hat{\rho}_{G}}\left[1 - \langle \alpha,\pm|\hat{\rho}_{G}|\alpha,\pm\rangle\right]}.
\end{equation}
To provide the same information as catability, it is normalized with the least infidelity attainable by Gaussian states. The normalized infidelity ranges from 0 for ideal cats to 1 for states with the same fidelity reachable by Gaussian approximations. It can, therefore, also serve as a witness of non-Gaussian behavior. Similarly to catability, we can construct a global normalized infidelity as
\begin{equation}\label{}
    \zeta = \min_{\alpha,\pm} \zeta_{(\pm)}(\alpha).
\end{equation}

We numerically evaluated the global catability and normalized infidelity for several cat states and their approximations subjected to loss. In particular, we considered even- and odd-parity cat states \eqref{eq:cat} with amplitudes ${\alpha \in \{ 0.5,1.5,2.5 \}}$. We also analyzed their photonic approximations \eqref{eq:singlephoton-approx} and \eqref{eq:even-approx}, with the tunable parameters set to ${r \approx - 5}$~dB and ${\omega = 0.618}$. These values minimize the expectation value of \eqref{eq:O_for_cat} for $\alpha = 1$ and can therefore be considered optimal approximation of small cat states.

\begin{figure}[ht!]
    \begin{center}
        \includegraphics[width=8.6cm]{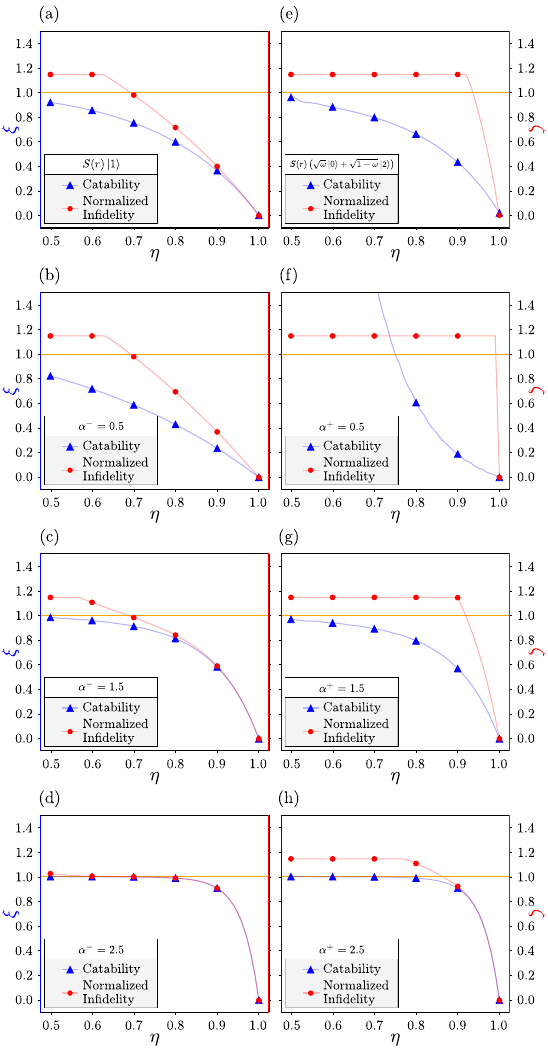}
    \end{center}\unskip
    \caption{
        Comparison of cat states subjected to different degrees of pure loss. The red curves represent the optimal values of the normalized infidelity \eqref{eq:normed_infi}, while the blue curves correspond to the optimal values of catability \eqref{eq:catability}, all relative to the transmissivity $\eta$ of the loss channel. The left column shows results for odd-parity cat states and their approximations, with \textbf{(a)}~squeezed single photon (${r \approx -5}$~dB), and cats with amplitudes ${\alpha^{-} \in \{ 0.5, 1.5, 2.5 \}}$ in panels \textbf{(b)}~through \textbf{(d)}, respectively. The results for even-parity states are presented in the right column, with \textbf{(e)} squeezed superposition of $\ket{0}$ and $\ket{2}$ (${\omega = 0.618}$, ${r \approx -5}$~dB), and cats with amplitudes ${\alpha^{+} \in \{ 0.5, 1.5, 2.5 \}}$ within panels \textbf{(f)} through \textbf{(h)}.
    }
    \label{fig:results}
\end{figure}

The results of the numerical simulations are presented in Figure \ref{fig:results}. The results for odd-parity states are organized in the left column, whereas the right column represents even states. A key observation is that even at loss levels where the normalized infidelity exceeds ${\zeta > 1}$, rendering the value inconclusive, catability successfully identifies cat states. Its sensitivity is strongest for small coherent states with amplitudes $0.5$ and $1$ and the small amplitude approximations \eqref{eq:singlephoton-approx} and \eqref{eq:even-approx}. We can also see that odd cat states retain their non-Gaussianity for a greater loss; the Wigner function of even-parity cats is positive at the origin and, for small amplitudes quickly, decoheres into a form that is difficult to distinguish from a Gaussian squeezed state.  Note that there is a fundamental difference in evaluation between even and odd cat states. Since the catability is normalized against Gaussian states that always have positive parity, large $\gamma$ in the optimization \eqref{eq:catability} can attribute significant non-Gaussian features even to states dissimilar from cat states. These states are still highly non-Gaussian, and the catability correctly picks up on this property. A possible way to compensate for this behavior is to choose a different class of states for normalization, such as photon-subtracted squeezed states (see \cite{SM} for details about optimization of $\alpha$ and $\gamma$ for odd cats). In the limit of low loss, catability aligns with fidelity for larger cat states and can be seen as a way to measure fidelity with lower resources.

The most significant advantage of catability lies in the direct measurability of the operator \eqref{eq:O_for_cat}, which can be decomposed into terms comprising number operators ${\hat{n} = \hat{a}^{\dagger}\hat{a}}$ transformed by Gaussian displacement operations ${\hat{D}(\beta) = \exp[ \beta^{\star} \hat{a} - \beta \hat{a}^{\dagger} ]}$ as
\begin{align}\label{eq:measurable_<O>}
    \begin{split}
        \hat{\mathcal{O}}_{(\pm)}(\alpha,\gamma) &=  2\hat{n}^2 + |\alpha|^2\left( 4\hat{n} +1\right) + 2|\alpha|^4 - \hat{n}  \\
        &- \frac{1}{2}\left[\hat{D}^{\dagger}(\alpha) \hat{n}^2 \hat{D}(\alpha) + \hat{D}(\alpha) \hat{n}^2 \hat{D}^{\dagger}(\alpha)\right]\\
        & + \gamma\left( 1\mp \hat{\Pi} \right) .
    \end{split}
\end{align}
The measurement statistics of the three constituent operators, $\hat{n}$, $\hat{D}^{\dagger}(\alpha) \hat{n}^{2} \hat{D}(\alpha)$, and $\hat{D}^{\dagger}(-\alpha) \hat{n}^{2} \hat{D}(-\alpha)$, are functions of the probabilities ${p_n(\beta) = \mathrm{Tr}[\hat{D}^{\dagger}(\beta) |n\rangle\langle n| \hat{D}(\beta) \hat{\rho}]}$ for any state $\hat{\rho}$ and can be used to determine the mean value of the operator \eqref{eq:measurable_<O>} as
\begin{align}\label{eq:measurable_<O>_pn}
    \begin{split}
        \langle \mathcal{\hat{O}}_{(\pm)}(\alpha,\gamma) \rangle 
        &= \sum_n p_n(0) \left[ 2n^2 - (1 - 4|\alpha|^2)n \right] \\
        &\mp \sum_n p_n(0) \left[ \gamma(-1)^n \right] \\
        &- \frac{1}{2} \sum_n p_n(\alpha)n^2 - \frac{1}{2}  \sum_np_n(-\alpha)n^2\\
        &+ 2|\alpha|^4 + |\alpha|^2 + \gamma.
    \end{split}
\end{align}
A benefit of this decomposition is that the measurement of the number operator is a feasible operation in all the contemporary platforms where the cat states are prepared \cite{endo2025,he2023,kienzler2016}. Therefore, the three sets of measurements are sufficient to verify the existence of the cat state.

To validate the feasibility of the direct measurement, we conducted a numerical simulation designed to closely resemble an actual experiment \cite{SM}. The simulation focuses on an odd-parity cat state with initial amplitude ${\alpha = 2}$ subjected to different amounts of amplitude loss, specifically to $0\%$, $10\%$, $20\%$, and $30\%$.

\begin{figure}[ht!]
    \begin{center}
        \includegraphics[width=8.6cm]{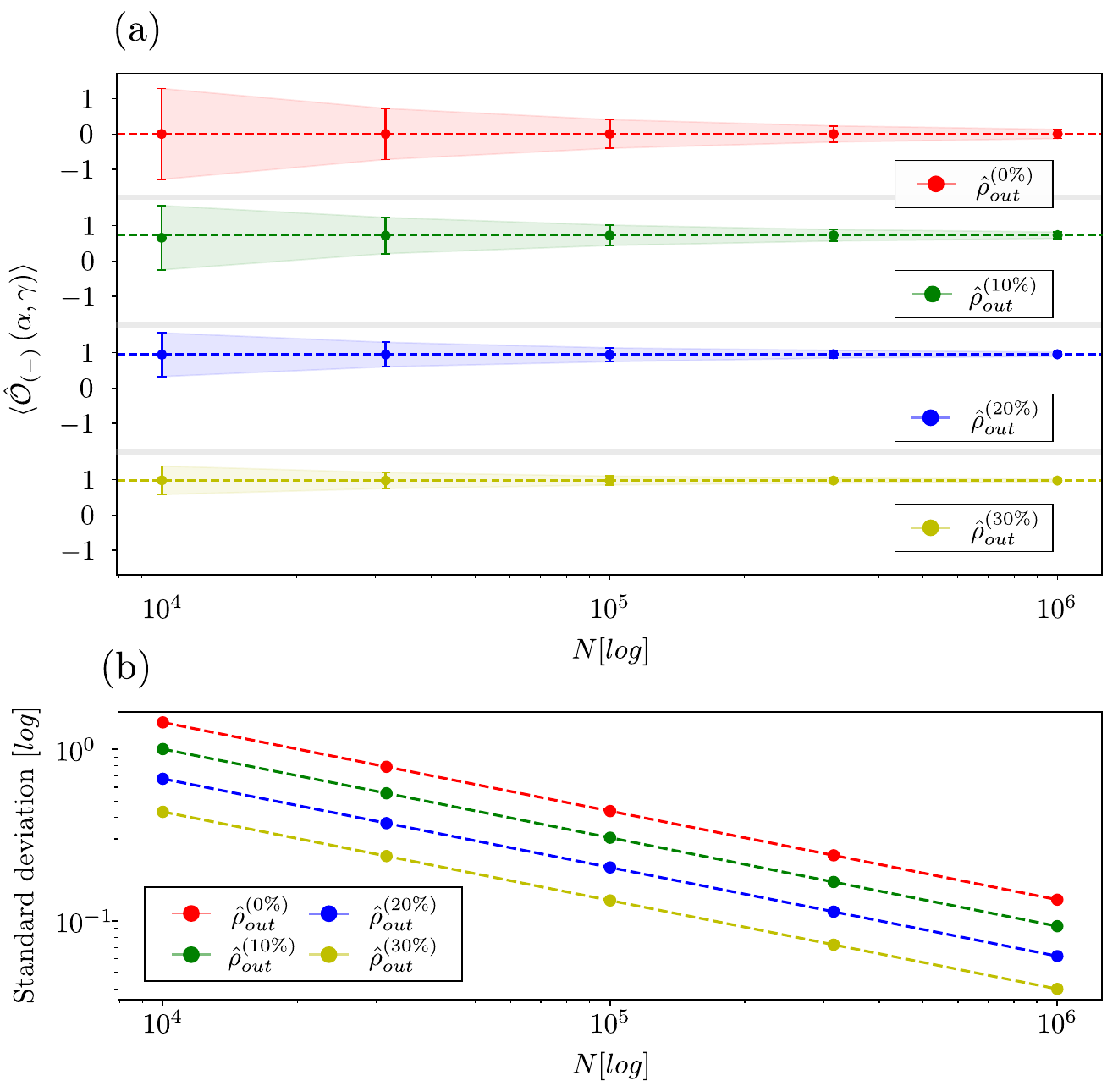}
    \end{center}\unskip
    \caption{
        Simulated direct measurement of the operator \eqref{eq:measurable_<O>_pn} for an odd-parity cat state with amplitude $\alpha = 2$ subjected to different levels of amplitude loss. Colors indicate $0\%$ (red), $10\%$ (green), $20\%$ (blue), and $30\%$ (yellow) loss.
        \textbf{(a)}~The dashed line indicates the mean value of the operator with respect to the number of performed measurements. The shaded regions represent uncertainty up to one standard deviation. 
        \textbf{(b)}~The standard deviation depends on the number of measurements. 
    }
    \label{fig:standard deviation}
\end{figure}

As shown in Figure \ref{fig:standard deviation}(a), the results indicate that the uncertainty consistently decreases as the number of measurements increases. The standard deviation is explicitly shown in Figure \ref{fig:standard deviation}(b). It is the largest for the state with zero loss, a consequence of the largest effective amplitude of the initial cat state.

As a final remark, note that while the measurement of catability for a fixed $\alpha$ requires only three sets of measurements of displaced number operators, which is not as difficult as the full tomography typically needed for fidelity~\cite{yukawa2013,eaton2022,endo2025}. The global catability demands scanning over all possible values of $\alpha$. When its value is known or can be coarsely estimated, for example, from the mean value of the number operator \cite{Ourjoumtsev2007Aug,Laghaout2015Feb}, our criterion offers a practical advantage over fidelity.

On a final note, let us discuss possible generalizations. Every balanced superposition of any two coherent states can be transformed into the form \eqref{eq:cat} by a displacement operation. Superpositions of displaced squeezed states can be accommodated by squeezing the operator \eqref{eq:O_for_cat} and instead employing 
\begin{equation}\label{eq:O_squeezed}
    \begin{aligned}
        \hat{\mathcal{O}}_{(\pm)}(\alpha,\gamma,r)
        = \; &\hat{S}^{\dagger}(r)\hat{\mathcal{O}}_{(\pm)}(\alpha,\gamma)\hat{S}(r) \\
        = \; & \left( (\mu \hat{a}^{\dagger} - \nu \hat{a})^{2} - \alpha^{*2} \right) \\
        & \cdot \left( (\mu\hat{a}-\nu\hat{a}^{\dagger})^{2} - \alpha^{2} \right) \\
        & + \gamma\left( 1 \mp \hat{\Pi} \right) ,\nonumber
    \end{aligned} 
\end{equation}
where ${\mu = \cosh(r)}$ and ${\nu = \sinh(r)}$. 

Another possible generalization employs superpositions of more coherent states, resulting in the multi-headed cat states
\begin{equation}
    \ket{\alpha,N,m} \propto \sum\limits_{k = 0}^{N-1} e^{\frac{ikm2\pi}{N}} \Ket{ \alpha e^{\frac{ik2\pi}{N}} },
\end{equation}
where $N$ indicates the number of its heads. These states are the ground eigenstates of the adapted operator
\begin{align}
    \begin{aligned}
                \hat{\mathcal{O}}^{(N)}(\alpha,\gamma,m)
                &= \left( \hat{a}^{\dagger N} - \alpha^{*N} \right)\left( \hat{a}^{N} - \alpha^{N} \right) \\
                &+ \gamma \left( 1- \sum_{k=1}^\infty \ket{Nk-m}\bra{Nk-m} \right),
    \end{aligned}
\end{align}
where ${m = 1,2,\dotsc,N}$ characterizes the particular phase factors between the constituent coherent states. Applications of these generalizations follow straightforwardly from the discussion of the two-headed cat scenario. In the supplementary material \cite{SM}, we demonstrate evaluation of catability for the three-headed cat state across all symmetries and compare the results with the corresponding normalized infidelity.

\emph{In conclusion}, we have introduced a new framework for identifying and characterizing cat states using an indicator based on the nonlinear squeezing of an operator whose ground states are ideal cat states. This method provides a reliable and experimentally accessible alternative to fidelity, which is often limited by its interpretability and dependence on full-state reconstruction. If needed, the nonlinear operator of catability can also be used to find optimal approximations of cat states constructed from states with limited stellar rank~\cite{chabaud2020,chabaud2021,fiurasek2022}.

Our results show that catability remains robust despite loss, efficiently distinguishing between ideal, approximate, and loss-affected cat states. It also offers the advantage of direct measurability through particle number distributions and displaced quadrature moments, enabling practical implementation in experimental setups.

Moreover, we have generalized the operator to support $N$-headed cat states, underlining the flexibility of the approach. As a criterion based on nonlinear squeezing, catability contributes a powerful tool for analysis of non-Gaussian quantum resources and opens new possibilities for state certification in realistic experiments~\cite{jezek2012,yukawa2013,eaton2022,endo2025,vlastakis2013,ofek2016,he2023, monroe1996,kienzler2016,omran2019,hacker2019,shomroni2020, Cohen2017Aug, hauer2023}.

\emph{Funding} We acknowledge support of the Czech Science Foundation (project 25-17472S), and European Union's HORIZON Research and Innovation Actions under Grant Agreement no. 101080173 (CLUSTEC). JP further acknowledges the Grant Agreements no. 101017733 and 731473 (CLUSSTAR). 
P.M. acknowledges the Programme Johannes Amos Comenius under the Ministry of Education, Youth and Sports of the Czech Republic reg. no. CZ.02.01.01/00/22\_008/0004649. VK acknowledges IGA\_PrF\_2025\_010. We also thank Michal Matulík and Jaromír Fiurášek for the valuable discussions.
We acknowledge use of the computational cluster at the Department of Optics at Palacký University and use of several open-source software libraries \cite{harris2020,virtanen2020,hunter2007} in the computation, evaluation and presentation of the results.

\emph{Data availability} The datasets supporting the presented results, along with the source code used for their creation, are publicly available \cite{source}.

\bibliography{manuscript.bib}

\clearpage
\onecolumngrid

\setcounter{section}{0}
\setcounter{figure}{0}
\setcounter{table}{0}
\renewcommand{\thesection}{S\arabic{section}}
\renewcommand{\thefigure}{S\arabic{figure}}
\renewcommand{\thetable}{S\arabic{table}}

\begin{center}
{\large\bfseries Supplementary Material for:}\\[1em]
{\Large Catability as a metric for evaluating superposed coherent states}\\[1em]
{\normalsize
\v{S}imon Br\"{a}uer, Jan Provazn\'{i}k, Vojt\v{e}ch Kala, Petr Marek\\
Optics Department, Faculty of Science, Palack\'{y} University\\
17. Listopadu 12, 77146 Olomouc, Czech Republic\\
(\today)
}
\end{center}

\vspace{1em}

\section{Gaussian Benchmark for the Operator $\hat{\mathcal{O}}$}

To determine the minimum expectation value of the operator $\hat{\mathcal{O}}$ over all Gaussian states, it is sufficient to consider only \textbf{pure Gaussian states}. This follows from the linearity of expectation values and the convexity of mixed states: any expectation value for a mixed state is a convex combination of those for pure states.

A general single-mode Gaussian state is fully characterized by:
\begin{itemize}
    \item Its \textbf{first moments} (mean quadrature values),
    \begin{equation} \label{eq:meanval}
    \epsilon = \begin{pmatrix}
    u \\
    v
    \end{pmatrix},
    \end{equation}
    \item and its \textbf{covariance matrix},
    \begin{equation} \label{eq:covmat}
    \sigma = 
    \begin{pmatrix}
    A & C \\
    C & B
    \end{pmatrix}.
    \end{equation}
\end{itemize}

We consider the operator
\begin{equation}
\hat{\mathcal{O}} = (\hat{a}^{\dagger 2} - \beta^*)(\hat{a}^2 - \beta) + \gamma (1 \pm \hat{\Pi}),
\end{equation}
where $\hat{a}$ is the annihilation operator, $\beta \in \mathbb{C}$ is a complex parameter, and $\hat{\Pi}$ is the parity operator.

Using the quadrature representation
\[
\hat{a} = \frac{1}{\sqrt{2}}(\hat{x} + i \hat{p}),
\]
the operator becomes
\begin{equation}
\begin{split}
\hat{\mathcal{O}} =& \frac{1}{4}\left(\hat{x}^4 + \hat{x}^2\hat{p}^2 + \hat{p}^2\hat{x}^2 + \hat{p}^4\right) - \hat{x}^2 - \hat{p}^2 \\
& - \text{Re}(\beta)(\hat{x}^2 - \hat{p}^2) - \text{Im}(\beta)(\hat{x}\hat{p} + \hat{p}\hat{x}) \\
& + |\beta|^2 + \gamma (1 \pm \hat{\Pi}) + \frac{3}{4}.
\end{split}
\end{equation}

To evaluate the expectation value of $\hat{\mathcal{O}}$ for a Gaussian state, we use moments calculated via the Wigner function formalism. In this representation, quantum moments of symmetrically ordered (Weyl-ordered) operators correspond directly to classical phase-space moments, evaluated over the Wigner function.

The Weyl ordering (also known as symmetric ordering) of an operator monomial $\hat{x}^n \hat{p}^m$ is defined as the fully symmetrized average over all possible orderings of $n$ position operators $\hat{x}$ and $m$ momentum operators $\hat{p}$. That is,
\[
:\!\hat{x}^n \hat{p}^m\!:_W = \frac{1}{(n+m)!} \sum_{\mathcal{P} \in S_{n+m}} \mathcal{P}\big(\hat{x}^{n} \otimes \hat{p}^{m}\big),
\]
where $S_{n+m}$ denotes the set of all permutations of the $n+m$ operators.

In particular, for any such monomial, the expectation value of the Weyl-ordered operator is given by the classical moment:
\[
\langle :\!\hat{x}^n \hat{p}^m\!:\rangle_W = \int dx\,dp\, W(x,p)\, x^n p^m.
\]
For Gaussian states, this quantity corresponds to a moment of a bivariate Gaussian distribution that can be explicitly constructed from \eqref{eq:meanval} and \eqref{eq:covmat}.

Moreover, the expectation value of the \textbf{parity operator} is directly related to the Wigner function at the origin:
\begin{equation}
\langle \hat{\Pi} \rangle = \pi W(0,0).
\end{equation}

Combining all relevant moments, the expectation value of $\hat{\mathcal{O}}$ is:
\begin{equation}
\begin{split}
\langle \hat{\mathcal{O}} \rangle = & \frac{1}{4}\big(v^4 + 2((u^2+A)v^2 + 4Cuv + Bu^2 + 2C^2 + AB) + 6Bv^2 \\
& + u^4 + 6Au^2 + 3B^2 + 3A^2 - 1\big) - u^2 - v^2 \\
& - \text{Re}(\beta)(u^2 - v^2 - B + A) - 2 \text{Im}(\beta)(uv + C) \\
& + |\beta|^2 - A - B + \frac{3}{4} \\
& + \gamma \left(1 \mp \frac{1}{2\sqrt{AB - C^2}} \exp\left[\frac{1}{2} \cdot \frac{u(Cv - Bu) + v(Cu - Av)}{AB - C^2}\right] \right).
\end{split}
\end{equation}

\subsection*{Covariance Matrix Parametrization}

To reduce the number of variational parameters, we reparametrize the covariance matrix $\sigma$ using \textbf{squeezing} and \textbf{rotation} parameters $(r, \theta)$. The vacuum covariance matrix
\[
\sigma_{\text{vac}} = \frac{1}{2} \mathbb{I}
\]
is transformed as:
\begin{equation}
\sigma = R(\theta) S(r) \sigma_{\text{vac}} S(r)^\top R(\theta)^\top,
\end{equation}
with the squeezing matrix
\begin{equation}
S(r) = 
\begin{pmatrix}
e^{-2r} & 0 \\
0 & e^{2r}
\end{pmatrix},
\end{equation}
and the rotation matrix
\begin{equation}
R(\theta) = 
\begin{pmatrix}
\cos\theta & \sin\theta \\
-\sin\theta & \cos\theta
\end{pmatrix}.
\end{equation}

This yields:
\begin{equation}
\begin{split}
A &= \frac{1}{2} e^{-4r} \cos^2\theta + \frac{1}{2} e^{4r} \sin^2\theta, \\
B &= \frac{1}{2} e^{4r} \cos^2\theta + \frac{1}{2} e^{-4r} \sin^2\theta, \\
C &= \frac{1}{2}(e^{4r} - e^{-4r}) \cos\theta \sin\theta.
\end{split}
\end{equation}

Next, we compute:
\[
AB - C^2.
\]

\paragraph{Step 1: Compute \(AB\)}
\begin{align*}
AB &= \left( \frac{1}{2} e^{-4r} \cos^2 \theta + \frac{1}{2} e^{4r} \sin^2 \theta \right) \left( \frac{1}{2} e^{4r} \cos^2 \theta + \frac{1}{2} e^{-4r} \sin^2 \theta \right) \\
&= \frac{1}{4} \left( \cos^4 \theta + \sin^4 \theta + (e^{8r} + e^{-8r}) \cos^2 \theta \sin^2 \theta \right).
\end{align*}

\paragraph{Step 2: Compute \(C^2\)}
\begin{align*}
C^2 &= \left( \frac{1}{2} (e^{4r} - e^{-4r}) \cos \theta \sin \theta \right)^2 \\
&= \frac{1}{4} (e^{8r} - 2 + e^{-8r}) \cos^2 \theta \sin^2 \theta.
\end{align*}

\paragraph{Step 3: Subtract}
\begin{align*}
AB - C^2 &= \frac{1}{4} \left( \cos^4 \theta + \sin^4 \theta + 2 \cos^2 \theta \sin^2 \theta \right) \\
&= \frac{1}{4} \left( 1 \right) = \frac{1}{4},
\end{align*}
where we used the identity:
\[
\cos^4 \theta + \sin^4 \theta = 1 - 2 \cos^2 \theta \sin^2 \theta.
\]

\subsection*{Simplified Expression}

Inserting this result into the previous formula, the expectation value simplifies to:
\begin{equation}
\begin{split}
\langle \hat{\mathcal{O}} \rangle = & \frac{1}{4}\big(v^4 + 2((u^2+A)v^2 + 4Cuv + Bu^2 + 2C^2 + \tfrac{1}{4}) + 6Bv^2 \\
& + u^4 + 6Au^2 + 3B^2 + 3A^2 - 1\big) - u^2 - v^2 \\
& - \text{Re}(\beta)(u^2 - v^2 - B + A) - 2 \text{Im}(\beta)(uv + C) \\
& + |\beta|^2 - A - B + \frac{3}{4} \\
& + \gamma \left(1 \mp \exp\left[2\left( u(Cv - Bu) + v(Cu - Av)\right)\right] \right).
\end{split}
\end{equation}
In the case of finding the extremum for the mean value $\langle \hat{\mathcal{O}} \rangle$, the optimization is done only through the parameters $r$ and $\theta$, which are contained in the terms of the covariance matrix $A$, $B$, and $C$. This reduces the difficulty of the problem by one dimension.

\section{Exploring intervals of $\alpha$ and $\gamma$ in catability optimization.}

In the main text, we discuss the optimization of catability and explore the roles of the parameters $\alpha$ and $\gamma$. Here, we aim to illustrate the individual contributions to the catability function: specifically, the expectation value of the operator $\hat{\mathcal{O}}$ in the test state (under varying loss conditions) and the corresponding minimal expectation over all Gaussian states. We now describe the optimization process in detail, using a concrete example based on a squeezed Fock state as the test state.

The test state is defined as a squeezed single-photon Fock state,
\begin{equation}
|\psi\rangle = \hat{S}(r) |1\rangle,
\end{equation}
with the squeezing operator $\hat{S}(r)$ and squeezing parameter $r = \frac{1}{4}$, corresponding to approximately $-5\,\mathrm{dB}$ of squeezing. The density matrix is then given by $\hat{\rho} = |\psi\rangle \langle \psi|$. To simulate photon losses, we apply a lossy channel described by the Kraus operators
\begin{equation}
\hat{M}_k = \sqrt{ \frac{(1 - \eta^2)^k}{k!} } \, \eta^{\hat{a}^\dagger \hat{a}} \, \hat{a}^k,
\end{equation}
where $\eta$ denotes the amplitude transmissivity ($\eta=1$ corresponds to no losses, $\eta=0.5$ to 50\% loss). The resulting lossy test state is
\begin{equation}
\hat{\rho}^{(\mathrm{test})} = \sum_{k=0}^{\infty} \hat{M}_k \hat{\rho} \hat{M}_k^\dagger.
\end{equation}

\begin{figure}[ht!]
    \centering
    \includegraphics[width=10cm]{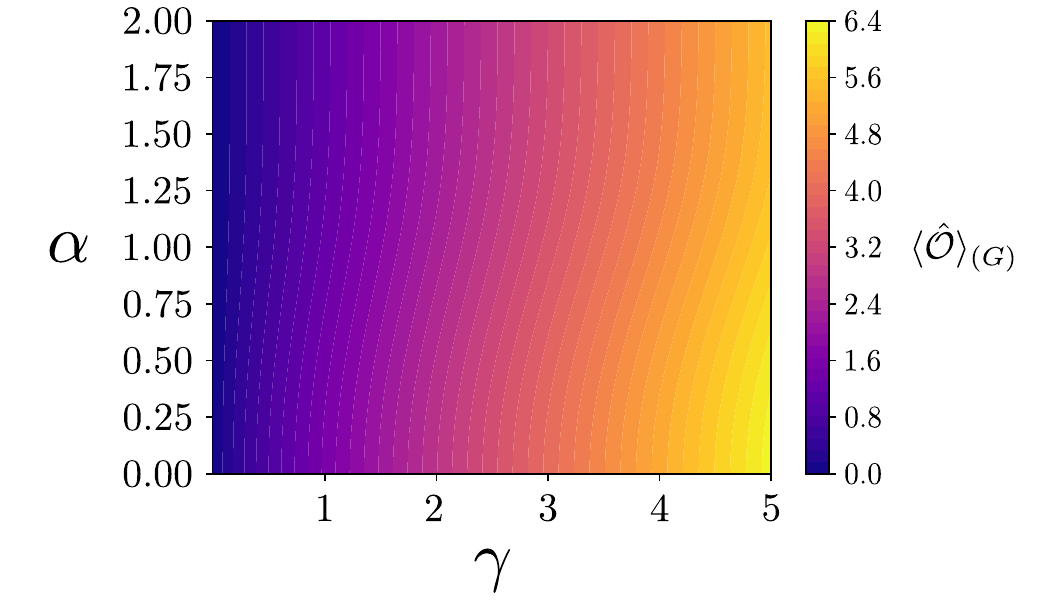}
    \caption{Display of optimal values of $\langle \hat{\mathcal{O}} \rangle_{(G)}$ \eqref{eq:meanvaluegauss}, minimized over all Gaussian states, in the studied interval ($\alpha$,$\gamma$).}
    \label{fig:gauss}
\end{figure}

We begin the optimization by evaluating the minimal expectation value of $\hat{\mathcal{O}}$ over all Gaussian states, i.e.,
\begin{equation}\label{eq:meanvaluegauss}
\langle \hat{\mathcal{O}} \rangle_{(G)} = \min_{\hat{\rho}^{(G)}} \text{Tr}\left( \hat{\rho}^{(G)} \hat{\mathcal{O}}_{(-)}(\alpha,\gamma)\right),
\end{equation}
where the minimization is performed over all Gaussian states in the specified parameter space. As shown in Fig.\ref{fig:gauss}, the minimal expectation values are achieved in a narrow region of $\gamma$, while spanning a broad range of $\alpha$. This reflects the fact that Gaussian states cannot exhibit negative parity.
We then evaluate the expectation value of $\hat{\mathcal{O}}$ for the test state across the same $(\alpha, \gamma)$ parameter space, under different levels of loss
\begin{equation}\label{eq:meanvaltest}
    \langle\hat{\mathcal{O}}\rangle_{(test)} =  \text{Tr}\left( \hat{\rho}^{(\text{test})} \hat{\mathcal{O}}_{(-)}(\alpha,\gamma)\right) 
\end{equation}

\begin{figure}[ht!]
    \centering
    \includegraphics[width=12cm]{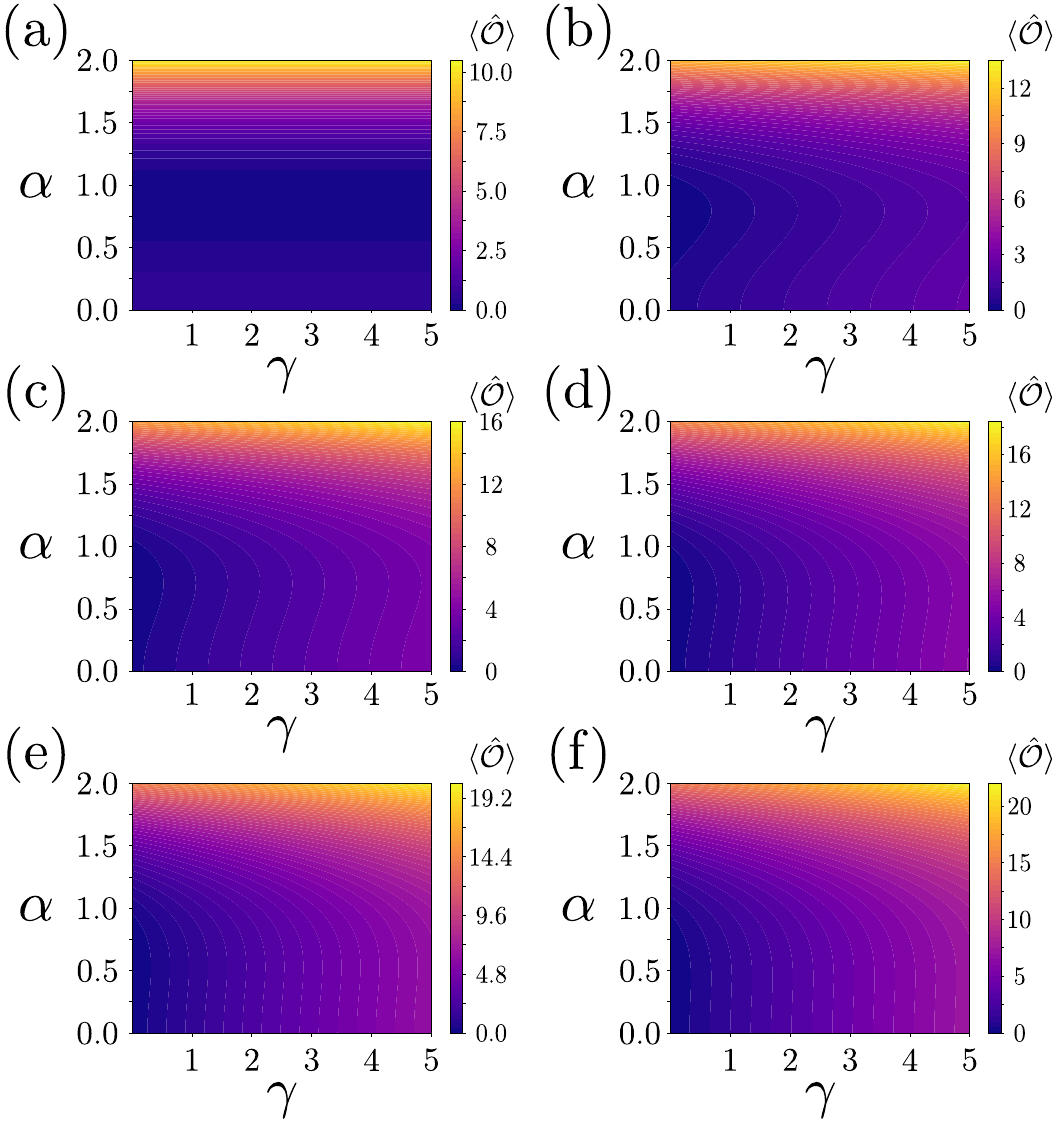}
    \caption{Display of optimal values of $\langle \hat{\mathcal{O}} \rangle_{(test)}$ \eqref{eq:meanvaltest}, in the studied interval ($\alpha$,$\gamma$). Individual figures show the input test state $\hat{\rho}^{(test)}$ (squeezed singlephoton state) with a different loss rate - (a) $\eta = 0$, (b) $\eta = 0.1$, (c) $\eta = 0.2$, (d) $\eta = 0.3$, (e) $\eta = 0.4$, (f) $\eta = 0.5$}
    \label{fig:testpotencial}
\end{figure}

The results are shown in Fig.~\ref{fig:testpotencial}, with subplots corresponding to loss levels from 0\% to 50\% (Fig.~\ref{fig:testpotencial}(a)--(f)). In the ideal case of a lossless, compressed single-photon state, Fig. \ref{fig:testpotencial}(a), the region of minimal mean value is primarily aligned along the x-axis (associated with the $\gamma$ parameter). This behavior is due to the well-defined parity of the state. In contrast, Gaussian states exhibit a different trend: the region of optimal values lies along the y-axis (associated with the $\alpha$ parameter), and as $\gamma$ increases, the mean value also increases—indicating degraded performance. This difference stems from the nature of Gaussian states, which have even (but not odd) parity at most. A similar effect can be observed when the test state undergoes losses: as losses increase, the state becomes more sensitive to the $\gamma$ parameter. This is due to the gradual Gaussification of the state resulting from the losses.

Note that catability below one can be, for some large values of gamma, reached also for mixtures of certain Fock states. In this case, catability detects the cat-like interference fringes at the origin of the phase space and confirms the non-Gaussian nature of the state, similar to the negativity of the Wigner function.

\section{Spectrum of the operator $\hat{O}_{(\pm)}$}

We consider the operator
\begin{equation}
\hat{\mathcal{O}}_{(\pm)}(\alpha,\gamma)
=\left(\hat{a}^{\dagger 2}-\alpha^{*2}\right)\left(\hat{a}^{2}-\alpha^{2}\right)
+\gamma\left(1\mp\hat{\Pi}\right),
\qquad
\hat{\Pi}=e^{i\pi\hat{a}^\dagger \hat{a}},
\end{equation}
where $\hat{a}$ and $\hat{a}^\dagger$ are the annihilation and creation operators, and $\hat{\Pi}$ denotes the photon-number parity operator.

\begin{figure}[ht!]
    \centering
    \includegraphics[width=18cm]{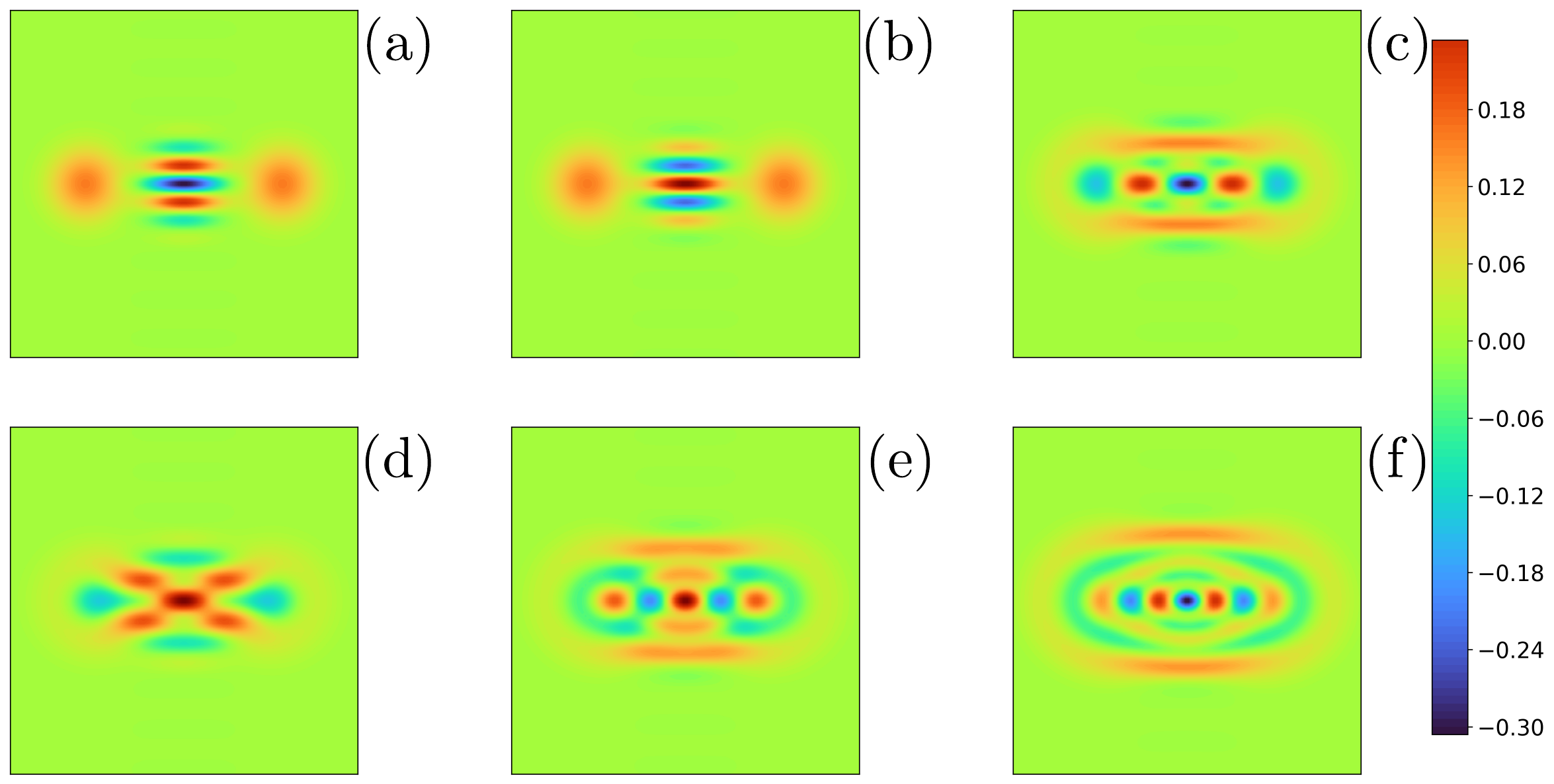}
    \caption{
        Wigner functions of the eigenstates of the operator $\hat{O}_{-}(\alpha = 2, \gamma = 1)$.  (a) Ground state of $\hat{O}_{-}$ corresponding to the \emph{odd} cat state.  (b) First excited eigenstate (non-ground state) corresponding to the \emph{even} cat state.  (c)–(f) Higher eigenstates of $\hat{O}_{-}(\alpha = 2, \gamma = 1)$, whose phase-space structures deviate from the ideal cat-state form. }
    \label{fig:eigenstate}
\end{figure}

If we perform a spectral analysis of the operator $\hat{\mathcal{O}}_{(\pm)}$, we find that its lowest-lying eigenstates are the even and odd Schrödinger cat states,
\begin{equation}
|C_\alpha^{(s)}\rangle = \mathcal{N}_s\left(|\alpha\rangle + s|-\alpha\rangle\right), 
\qquad 
s=\pm 1,
\quad
\mathcal{N}_s = \frac{1}{\sqrt{2\left(1+s\,e^{-2|\alpha|^2}\right)}}.
\end{equation}
Since $(\hat{a}^2-\alpha^2)|\pm \alpha\rangle=0$, the quartic term in $\hat{\mathcal{O}}_{(\pm)}$ annihilates any state in $\mathrm{span}\{|\alpha\rangle,|-\alpha\rangle\}$. The operator $\hat{\Pi}$ acts as $\hat{\Pi}|C_\alpha^{(s)}\rangle = s|C_\alpha^{(s)}\rangle$. Therefore, the action of $\hat{\mathcal{O}}_{(\pm)}$ on the cat states gives:
\begin{equation}
\hat{\mathcal{O}}_{(\pm)}|C_\alpha^{(s)}\rangle
= \gamma(1\mp s)\,|C_\alpha^{(s)}\rangle.
\end{equation}
Hence, the two lowest eigenvalues are
\begin{equation}
\lambda_0 = 0 \quad \text{for the cat with matching parity } (s=\pm1), 
\qquad
\lambda_1 = 2\gamma \quad \text{for the opposite parity cat.}
\end{equation}
Higher eigenstates of $\hat{\mathcal{O}}_{(\pm)}$ do not resemble cat states and their corresponding eigenvalues are integer multiples of $\gamma$. For illustration, Fig.~\ref{fig:eigenstate} in the main text shows the Wigner functions of the first six eigenstates of $\hat{\mathcal{O}}_{(+)}(\alpha=2,\gamma=1)$, demonstrating this structure.

\subsection*{Expectation Value for Coherent (Gaussian) States}

For a coherent state $|\beta\rangle$, the operator $\hat{a}$ acts as $\hat{a}|\beta\rangle=\beta|\beta\rangle$, and $\hat{\Pi}|\beta\rangle=|-\beta\rangle$. Using these relations,
\begin{equation}
(\hat{a}^2-\alpha^2)|\beta\rangle=(\beta^2-\alpha^2)|\beta\rangle,
\qquad
\langle\beta|(\hat{a}^{\dagger 2}-\alpha^{*2})=(\beta^{*2}-\alpha^{*2})\langle\beta|.
\end{equation}
This yields the expectation value
\begin{equation}
\boxed{
\langle \hat{\mathcal{O}}_{(\pm)}(\alpha,\gamma)\rangle_\beta
=|\beta^2-\alpha^2|^2
+\gamma\left(1\mp e^{-2|\beta|^2}\right).
}
\end{equation}
For $\beta=\pm\alpha$, the quartic term vanishes, and we obtain
\begin{equation}
\langle \hat{\mathcal{O}}_{(\pm)}\rangle_{\beta=\pm\alpha}
=\gamma\left(1\mp e^{-2|\alpha|^2}\right)
\simeq \gamma, \quad (|\alpha|\gtrsim 2),
\end{equation}
i.e., for realistic cat amplitudes, the mean value for a Gaussian (coherent) state approaches $\gamma$.

\subsection*{Implications for the Catability Metric}

The spectral decomposition reveals that the ground state and first excited state of $\hat{\mathcal{O}}_{(\pm)}$ are the cat states of corresponding and opposite parity, with eigenvalues $\lambda_0=0$ and $\lambda_1=2\gamma$, respectively. Higher eigenstates deviate from cat-like profiles, with eigenvalues that are integer multiples of $\gamma$.

When employing this operator in the definition of the \emph{catability} metric, the expectation value is normalized by its value over Gaussian states. Considering a coherent state as the representative Gaussian state (which is valid for $|\alpha|\simeq2$), we obtain an expectation value equal to $\gamma$. Consequently, a Gaussian state always provides a lower mean value than a cat state with opposite parity.

In scenarios where the system is not in a pure cat state but in a mixture of an ideal cat state (of a given parity) and Gaussian noise, the corresponding expectation value of $\hat{\mathcal{O}}_{(\pm)}$ lies within the range
\begin{equation}
0 \le \langle \hat{\mathcal{O}}_{(\pm)} \rangle \le \gamma,
\end{equation}
where the upper bound is reached for Gaussian states. This behavior establishes a direct quantitative link between the spectral properties of $\hat{\mathcal{O}}_{(\pm)}$ and the catability metric.

\section{N-headed cat states}

To demonstrate the applicability of the catability measure for multi-headed cat states, we extend the formalism introduced in the main text to the case of a three-headed cat ($N=3$). The corresponding operator is given by
\begin{equation}
\hat{\mathcal O}^{(3)}(\alpha,\gamma,m)
= (\hat a^{\dagger 3} - \alpha^{*3})(\hat a^{3} - \alpha^{3})
+ \gamma \left( 1 - \sum_{k=1}^{\infty} |3k - m\rangle\langle 3k - m| \right),
\label{eq:O3}
\end{equation}
where $m \in \{0,1,2\}$ labels the rotational symmetry sector of the state.

The ground eigenstates of $\hat{\mathcal O}^{(3)}(\alpha,\gamma,m)$ are the three-headed cat states
\begin{equation}
|\alpha,3,m\rangle 
= \mathcal N_{3,m}(\alpha) 
\sum_{k=0}^{2} e^{i k m \frac{2\pi}{3}} 
\big|\alpha e^{i k \frac{2\pi}{3}}\big\rangle,
\label{eq:3cat}
\end{equation}
where $\mathcal N_{3,m}(\alpha)$ is the normalization factor ensuring 
$\langle \alpha,3,m | \alpha,3,m \rangle = 1$.

As in the main text, we perform numerical simulations to test the robustness of these states under optical loss. The lossy channel is modeled by amplitude transmittance $0 \le \eta \le 1$ through Kraus operators
\begin{equation}
\hat M_k = \sqrt{\frac{(1-\eta^2)^k}{k!}}\, \eta^{\hat a^{\dagger}\hat a} \hat a^k,
\label{eq:kraus}
\end{equation}
so that the output state is
\begin{equation}
\hat \rho_{\text{out}} = \sum_{k=0}^{\infty} \hat M_k \hat \rho \hat M_k^{\dagger}.
\label{eq:loss}
\end{equation}
We consider the case of $30\%$ photon loss, i.e., $\eta^2 = 0.7$, which strongly suppresses the non-Gaussian interference fringes of the ideal cat states.
\begin{figure}[ht!]
    \begin{center}
        \includegraphics[width=17.2cm]{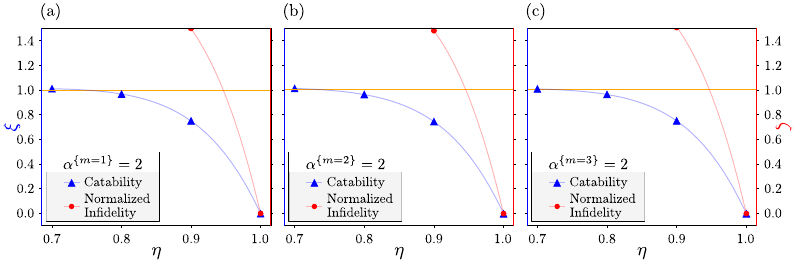}
    \end{center}\unskip
    \caption{
        Comparison of 3-headed cat states subjected to different degrees of pure loss. The red curves represent the optimal values of the normalized infidelity \eqref{eq:infidelity}, while the blue curves correspond to the optimal values of catability \eqref{eq:xi3}, all relative to the transmissivity $\eta$ of the loss channel. \textbf{(a)}~3-headed cat with amplitudes ${\alpha = 2\}}$ with symmetry $m=1$, \textbf{(b)}~3-headed cat with amplitudes ${\alpha = 2\}}$ with symmetry $m=2$ and \textbf{(c)}~3-headed cat with amplitudes ${\alpha = 2\}}$ with symmetry $m=3$.
    }
    \label{fig:N-headed}
\end{figure}

The catability of the resulting state $\hat \rho_{\text{out}}$ is then evaluated as
\begin{equation}
\xi_{(m)}(\alpha) 
= \min_{\gamma>0} 
\frac{\mathrm{Tr}\!\left[\hat{\mathcal O}^{(3)}(\alpha,\gamma,m)\hat \rho_{\text{out}}\right]}
{\displaystyle \min_{\hat \rho_G} 
\mathrm{Tr}\!\left[\hat{\mathcal O}^{(3)}(\alpha,\gamma,m)\hat \rho_G\right]},
\label{eq:xi3}
\end{equation}
where the minimization in the denominator is performed over all Gaussian states $\hat \rho_G$. The minimization over $\gamma$ ensures optimal sensitivity of the criterion for a given $\alpha$ and symmetry index $m$.

In Fig.\ref{fig:N-headed}, we show the evaluation of the three-headed catability for all three symmetry sectors ($m=0,1,2$) with the coherent-state amplitude fixed at $\alpha = 2$. For comparison, we also plot the normalized infidelity $\zeta_{(m)}(\alpha)$ defined as
\begin{equation}\label{eq:infidelity}
\zeta_{(m)}(\alpha) = 
\frac{1 - \langle \alpha,3,m | \hat \rho | \alpha,3,m \rangle}
{\displaystyle \min_{\hat \rho_G} \left[ 1 - \langle \alpha,3,m | \hat \rho_G | \alpha,3,m \rangle \right]},
\end{equation}
which quantifies the deviation from the target three-headed cat state relative to the best Gaussian approximation.  
In all three symmetry sectors, the infidelity exceeds the Gaussian threshold already at about $10\%$ of loss, implying that in this regime the measure becomes inconclusive about the non-Gaussian nature of the state.  
In contrast, the catability criterion remains below the threshold up to approximately $30\%$ of loss, indicating that it still reliably certifies the presence of cat-like coherence in moderately degraded states, even when the normalized infidelity no longer provides information.

\section{Measurement}
Most methods of state evaluation require a density matrix from which various indicators are then calculated. This means that a full-state tomography needs to be done. The advantage of catability is that the mean value of the operator $\hat{\mathcal{O}}$ can be measured directly:
    \begin{align}\label{eq:measurable_<O>}
        \begin{split}
            \langle\hat{\mathcal{O}}_{(\pm)}(\alpha,\gamma) \rangle &= \langle2\hat{n}^2 + |\alpha|^2\left( 4\hat{n} +1\right) + 2|\alpha|^4 \\
            &- \frac{1}{2}\left[\hat{D}^{\dagger}(\alpha) \hat{n}^2 \hat{D}(\alpha) + \hat{D}(\alpha) \hat{n}^2 \hat{D}^{\dagger}(\alpha)\right]\\
            &- \hat{n} + \gamma\left( 1\mp \hat{\Pi} \right) \rangle\\
            &= \sum_n p_n(0) \left[ 2n^2 - (1 - 4|\alpha|^2)n \mp \gamma(-1)^n \right] \\
            &-\frac{1}{2} \sum_n p_n(\alpha)n^2 -\frac{1}{2} \sum_np_n(-\alpha)n^2\\
            &+ 2|\alpha|^4 + |\alpha|^2 + \gamma,
        \end{split}
    \end{align}
where $p_n(\beta) = \mathrm{Tr}[\hat{D}^{\dagger}(\beta) |n\rangle\langle n| \hat{D}(\beta) \hat{\rho}^{(test)}]$ are probability distributions obtained by measuring $\hat{n}$ in a state displaced by coherent amplitude $\beta$.

We performed a Monte Carlo numerical simulation of the relevant experimental measurement. For each particular $\alpha$, the input was a density matrix of the tested quantum state, $\hat{\rho}^{(test)}$. For this state, we have considered three measurement setups resulting in three probability distributions, $p_n(0)$, $p_n(\alpha)$, and $p_n(-\alpha)$. For each scenario, we constructed a displaced density matrix $\hat{D}(\beta)\hat{\rho}^{(test)}\hat{D}^{\dagger}(\beta)$ from which we obtained the required probability distributions $p_n(\beta)$ simply by taking the diagonal.

We have then randomly generated three sets of real numbers occuring with these probabilities, each of length $N$. From them, we have calculated the experimental frequencies $f_n(\beta,N)$ and used them in (\ref{eq:measurable_<O>}), in places of $p_n(\beta,N)$, to obtain an estimate of catability. For each $N$, we have repeated the numerical experiment 1 000 000 times to obtain the mean values and standard deviations that are shown in the manuscript.

\section{Complete measurement protocol}
Using catability to verify a specific cat state is a straightforward task requiring only three sets of number operator measurement. Characterizing quality of a  completely unknown cat state is a more challenging task, albeit one that can still be streamlined. In such scenario, the full measurement protocol is separated into two steps. The goal of the first step is to estimate the form of the state, while the second step then focuses on characterizing the quality. 

The first step bears some similarity to estimation of the state. Measurements of displaced number operators reveal point values of the state's Wigner function. For evaluation of the catability, however, it is not necessary to obtain the full Wigner function with high precision, just the location of the main peaks of the cat, which reduces the number of required measurement runs. After the basic shape of the cat is recognized, the cat is oriented in the phase space, and characterization is performed in the second step by three displaced number operator measurements with large number of measurement runs, as detailed in the previous section. 

While the first step may require more different measurement settings, the second step may well be more demanding in the number of measurement runs. This is actually desirable, as the second step realizes the desired characterization. Furthermore, mistakes in the first step of the protocol can never lead to false positives. 

The first step of the protocol can also take advantage of any prior information about the state to reduce the number of required measurements. For example, if we know the produced state is a cat state centered at the origin of the phase space, the full characterization protocol may look like this:
\begin{itemize}
	\item[(1)] Measure $\hat{n}$ to obtain frequencies $f_n(0)$. Use them to estimate $|\alpha|$ of the state. This step consists of $N_0$ individual measurement runs 
	\item[(2)] Perform a set of measurements of operators $D^{\dag}(|\alpha|e^{i\phi})\hat{n}D(|\alpha|e^{i\phi})$ to obtain distribution $P(\phi)$, where $\phi \in [0,\pi)$. Identify the values $\phi_m$ in which $P(\phi)$  attains the maximal value.   
	\item[(3)] Measure displaced number operators $D^{\dag}(|\alpha|e^{i\phi_m})\hat{n}D(|\alpha|e^{i\phi_m})$ and  $D^{\dag}(-|\alpha|e^{i\phi_m})\hat{n}D(-|\alpha|e^{i\phi_m})$ to obtain the two remaining frequencies for (\ref{eq:measurable_<O>}). Each of these measurements consist of $N_0$ individual measurement runs. 
\end{itemize}

\end{document}